\documentclass[preprint,12pt,authoryear]{elsarticle}

\usepackage{graphicx}
\usepackage{bm}
\usepackage{mathptmx}
\usepackage[T1]{fontenc}
\usepackage{amsmath,amssymb}
\usepackage{hyperref}
\usepackage{subcaption}

\usepackage{lineno}
\modulolinenumbers[1]

\journal{Urban Forestry \& Urban Greening}

\begin{document}
\section*{Authors and affiliations}

\noindent
\textbf{T. Tokiwa}\\
Faculty of Engineering, Institute of Science Tokyo, Tokyo, 152-8550, Japan.\\
Email: \href{mailto:tokiwa.t.3c84@m.isct.ac.jp}{tokiwa.t.3c84@m.isct.ac.jp}

\vspace{3mm}
\noindent
\textbf{Y. Yin}\\
Faculty of Engineering, Institute of Science Tokyo, Tokyo, 152-8550, Japan.\\
Email: \href{mailto:yin.y.1b8b@m.isct.ac.jp}{yin.y.1b8b@m.isct.ac.jp}

\vspace{3mm}
\noindent
\textbf{R. Onishi}\\
Supercomputing Research Center, Institute of Integrated Research, Institute of Science Tokyo, Tokyo, 152-8550, Japan.\\
Email: \href{mailto:onishi.ryo@scrc.iir.isct.ac.jp}{onishi.ryo@scrc.iir.isct.ac.jp}

\vspace{6mm}
\section*{Corresponding author}
\noindent
T. Tokiwa (\href{mailto:tokiwa.t.3c84@m.isct.ac.jp}{tokiwa.t.3c84@m.isct.ac.jp})

\vspace{6mm}
\section*{Acknowledgements}
\noindent
This work used computational resources TSUBAME4.0 supercomputer provided by Institute of Science Tokyo through Joint Usage/Research Center for Interdisciplinary Large-scale Information Infrastructures and High Performance Computing Infrastructure in Japan (Project ID: jh240041).

\begin{frontmatter}

\title{Drag Crisis in Fractal Trees Revealed by Simulation and Theory}


\begin{abstract}
Trees are key roughness elements in urban environments, strongly influencing airflow, microclimates, and pollutant dispersion. While aerodynamic drag is central to these processes, its behavior for complex tree-like structures at high Reynolds numbers remains poorly characterized, in contrast to the well-documented drag crisis of simple bluff bodies. Here, we combine large-scale lattice Boltzmann simulations with an analytical branch-wise drag model to investigate the drag of fractal tree geometries over a wide range of tree-height-based Reynolds numbers, $Re_H$. Direct numerical simulations with a cumulant lattice Boltzmann method and adaptive mesh refinement are performed for $2.5\times 10^3 \le Re_H \le 1.2\times 10^5$, and the analytical model is then applied up to $Re_H \sim 10^9$. Under uniform inflow, the analysis shows that a distinct drag crisis is expected near $Re_H \approx 3\times 10^6$, but that the sharpness of this transition decreases with increasing structural complexity because smaller branches remain subcritical. A sensitivity test incorporating inflow turbulence with a streamwise turbulence intensity $I_u \approx 8\%$, representative of atmospheric-boundary-layer winds over rough urban surfaces, shifts the apparent onset of the crisis to $Re_H \approx 1.5\times 10^5$ and further smooths the drag reduction. Interpreted in terms of full-scale parameters, this implies that urban trees of order $10$--$30$~m height exposed to typical near-surface wind speeds of order $1$--$10~\mathrm{m/s}$ generally experience drag in the crisis or post-crisis regime. In both uniform and turbulent inflow cases, the framework further predicts that the relative ordering of drag coefficients across geometries reverses with $Re_H$: simplified trees experience lower drag in the subcritical regime but can exhibit higher drag in the supercritical regime, whereas more complex trees display a smoother, moderated crisis. This robustness challenges the common assumption in tree-management practice that pruning, which reduces structural complexity, always decreases aerodynamic loading, and highlights the need to reassess vegetation-drag parameterizations and pruning strategies in light of the relevant high-$Re_H$ regime.
\end{abstract}


\begin{keyword}
fractal tree, lattice Boltzmann method, adaptive mesh refinement,
L-system, drag coefficient, drag crisis
\end{keyword}

\end{frontmatter}


\section{Introduction}
In recent years, the need for accurate urban micrometeorology forecasts has grown increasingly urgent, driven by accelerating urban population growth~\citep{UN:WUP2019}, climate change~\citep{IPCC2023,UNHabitat2024}, and the rise of smart cities. Such need requires a detailed understanding of the physical processes that shape the urban atmosphere, including the effects of vegetation.

Trees play a vital role in regulating the micrometeorology of the urban environment by modulating wind flow, temperature distribution, and pollutant dispersion~\citep{Gromke2015a,Gromke2015b,Oshio2021,Reznicek2025}. These effects arise from complex aerodynamic interactions between airflow and trees.
As cities strive to enhance climate resilience and ensure public safety, a deep understanding of the aerodynamic behavior of trees becomes increasingly critical. Among various aerodynamic parameters, the drag characteristics of trees significantly influence wind-tree interactions, yet remain insufficiently understood.

In particular, the drag coefficient, a dimensionless parameter that quantifies the resistance a tree exerts on wind flow, plays a central role in determining the drag characteristics.
Previous experimental and numerical studies have examined the drag coefficient of trees in isolated cases; however, a comprehensive theoretical framework or a systematic understanding of how tree drag depends on flow conditions such as the Reynolds number ($Re$, a dimensionless parameter characterizing the ratio of inertial to viscous forces in a flow) has yet to be developed~\citep{Hagen1971,Grant1998,Gillies2002,Guan2003,Dong2008,Bitog2011,Manickathan2018,Wilson1985}. To our knowledge, only one numerical study~\citep{Yin2025}, which analyzed the total drag coefficient of trees, has provided a systematic understanding of $Re$ dependency for $Re$ up to $10^5$. However, that study could not extend into higher Reynolds number regime. Experimental investigations using wind tunnels are constrained by the scaling limitations of physical models~\citep{Ahmad2005,Lateb2016,Tominaga2013,Oke1988}, while high-fidelity numerical simulations demand substantial computational resources~\citep{Baik2002,Salim2011,Neofytou2008,Mei2019}. Furthermore, the phenomenon of drag crisis, a sharp drop in drag coefficient due to boundary layer transition to turbulence, commonly observed in simple bluff bodies~\citep{Singh2005,Smith1999,Choi2006}, has not been observed in trees because of the extremely high Reynolds number required.

To address these challenges, we employed a multi-GPU-based lattice Boltzmann method (LBM)~\citep{Geier2015}, which offers a more computationally efficient alternative to conventional multi-CPU simulations, to systematically evaluate the drag coefficient of trees. Notably, our analysis decomposes the total drag coefficient into its constituent components: friction drag and pressure drag. 
To generate realistic yet controllable tree models, we utilized the L-system employed in~\citep{Segrovets2022}, which enables systematic model construction. In this study, we analyze a type of fractal tree across three iterations: $n=4,6$ and 8, where larger $n$ generates more complex tree.

Despite the advantages of the multi-GPU-based LBM, simulating tree aerodynamics at very high Reynolds numbers remains computationally prohibitive. Therefore, to circumvent these limitations, we propose a novel analytical framework that enables estimation of drag coefficients, including the potential onset of drag crisis, in trees under high-$Re$ flow conditions. This analytical method takes advantage of the geometric characterization of tree models generated by the L-system, where we modeled the fractal tree as an assembly of cylindrical segments. This approach offers a promising pathway to understanding the complex interplay between tree morphology and aerodynamic response beyond the reach of current experimental and numerical methods.

The remainder of this paper is organized as follows.
In Sec.~II, we describe the fractal-tree geometries, the LBM-AMR numerical method, the simulation setup, and the analytical drag-estimation framework.
In Sec.~III, we present the LBM results and examine how the total and
component-wise drag coefficients depend on Reynolds number and tree
morphology, and validate the analytical framework against the simulations.
In Sec.~IV, we apply the analytical framework to estimate drag over an
extended Reynolds number range that is beyond the reach of our simulations
and identify the conditions under which a drag crisis is expected.
In Sec.~V, we discuss the implications of these findings for drag reversal
and urban tree management.
Finally, Sec.~VI summarizes the main conclusions and provides an outlook
for future work.

\section{Methods}
\subsection{Fractal tree models}
To generate fractal tree structures, we employed the L-system algorithm,~\citep{Prusinkiewicz1996}
a parametric, rule-driven approach capable of producing intricate branching
patterns from a compact set of input parameters. This method enables efficient
modeling of natural, self-similar forms through graphical interpretation.

\subsubsection{L-system}
The parametric L-system allows key characteristics such as the number of
branching generations, as well as the length, diameter, and angle of branches,
to be systematically defined. The underlying algorithm of the parametric
L-system operates as follows:
\begin{gather}
\begin{split}
\omega&:A\left(s_0,w_0\right),\\
p_1&:A\left(s,w\right):s\geq \min\rightarrow~ !\left(w\right)F\left(s\right)\\
&\quad\left[+\left(\alpha_1\right)/\left(\varphi_1\right)A\left(s\ast r_1,w\ast q^e\right)\right]\\
&\quad\left[+\left(\alpha_2\right)/\left(\varphi_2\right)A\left(s\ast r_2, w\ast (1-q)^e\right)\right],
\end{split}
\label{eq:L-system-algorithm}
\end{gather}
where the first line specifies the parameters associated with the tree trunk.
The symbol $A$ denotes an apex, corresponding to a not-yet-generated branch
(modeled as a cylinder) with normalized length $s_0$ and diameter $w_0$. The
subsequent lines define the branching rules, where $p_1$ represents the
transformation of the apex $A$ into an internode $F$ (i.e., a branch segment)
and two additional apices $A$.

\begin{figure}[htbp]
\centering
\includegraphics[scale=0.30]{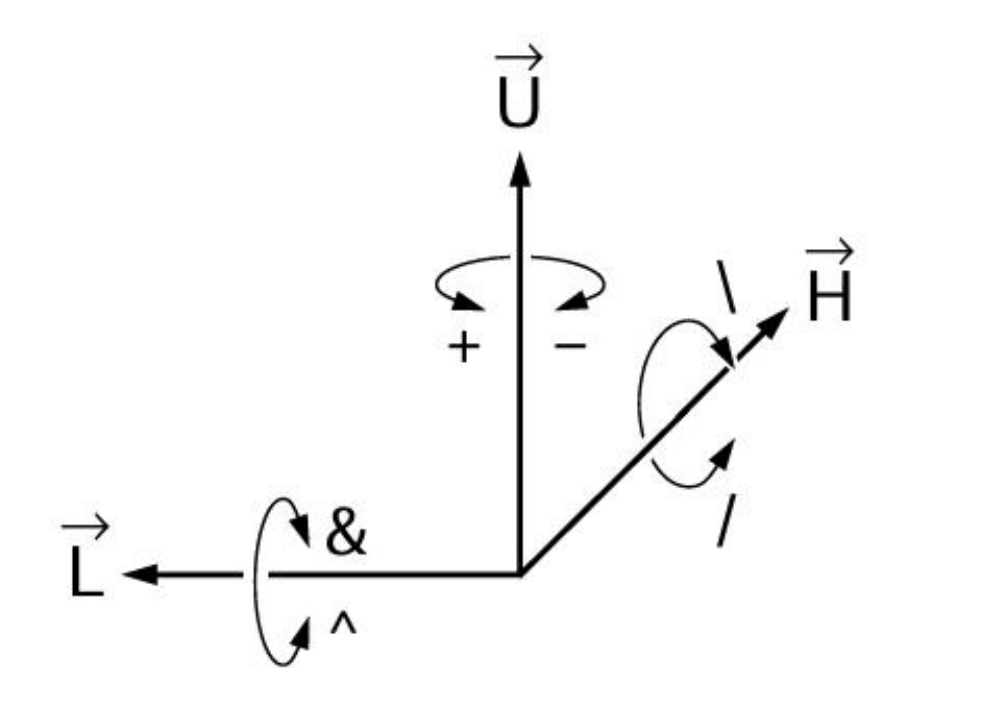}
\caption{Axis conventions $(+, -, \&, \wedge, \backslash, /)$ for the
L-system turtle interpretation. The vectors $\vec{H}$, $\vec{L}$, and
$\vec{U}$ denote the heading, left, and up directions, respectively.}
\label{fig:coordinates}
\end{figure}

When the length $s$ of an apex exceeds or equals a predefined minimum
threshold $\min$, it generates a branch according to the parameters $(s,w)$
assigned to that apex. Simultaneously, two new apices are produced, with their
parameters scaled by $s r_1$, $s r_2$, $w q^e$, and $w (1- q)^e$, respectively,
to define the dimensions of the next branches. The orientation of these new
apices relative to the preceding branch is governed by the rotation symbols
``$+$'' and ``/''. The axis conventions for these transformations are
illustrated in Fig.~\ref{fig:coordinates}, where $\vec{H}$ denotes the
direction of the preceding branch (i.e., the longitudinal axis of the
cylinder), while $\vec{L}$ and $\vec{U}$ represent the directions to the left
and upward, respectively. These three vectors are mutually orthogonal,
satisfying $\vec{H}\times \vec{L}=\vec{U}$.

The replacement rule $p_1$ is applied $n$ times to generate the desired number
of branches, after which the full three-dimensional geometry of the fractal
tree can be constructed. A summary of symbols used in
Eq.~\eqref{eq:L-system-algorithm} is given in
Table~\ref{tab:L-system-symbols}.

\begin{table}[htbp]
\caption{\label{tab:L-system-symbols}\centering
Summary of symbols used in the L-system formulation
in Eq.~\eqref{eq:L-system-algorithm}.}
\begin{tabular}{lp{0.78\textwidth}}
$A(s,w)$       & Apex; represents a not-yet-generated branch with parameters
                 $(s,w)$ and an implicit turtle state $\vec{P}$\\
$\vec{P}^i_j$  & Turtle state associated with apex $A$, specifying origin and
                 orientation via $\vec{U}$, $\vec{L}$, and $\vec{H}$\\
$\ast$         & Multiplication operator\\
$i$            & Index distinguishing different branch generations\\
$j$            & Index distinguishing between branches of the same generation\\
$F(s)$         & Branch segment of length $s$ generated when rule $p_1$ is
                 applied\\
$\min$         & Minimum threshold for branch length\\
$\omega$       & Axiom; initial symbol that begins the string rewriting\\
$s$            & Branch length\\
$w$            & Branch diameter\\
$\alpha,\varphi$ & Rotation angles\\
$+$, $-$       & Clockwise / counterclockwise rotation about $\vec{U}$\\
$\backslash$, / & Clockwise / counterclockwise rotation about $\vec{H}$\\
\&, $\wedge$   & Clockwise / counterclockwise rotation about $\vec{L}$\\
$!(w)$         & Sets the branch diameter to $w$\\
$[$            & Starts a new state $\vec{P}^{i+1}_j$; subsequent actions are
                 relative to $\vec{P}^{i+1}_j$\\
$]$            & Ends the operations related to $\vec{P}^{i+1}_j$ and returns
                 to the previous state\\
\end{tabular}
\end{table}

\subsubsection{Geometric structure and dimensional characteristics of fractal trees}

\begin{figure}[htbp]
\centering
\includegraphics[width=0.25\textwidth]{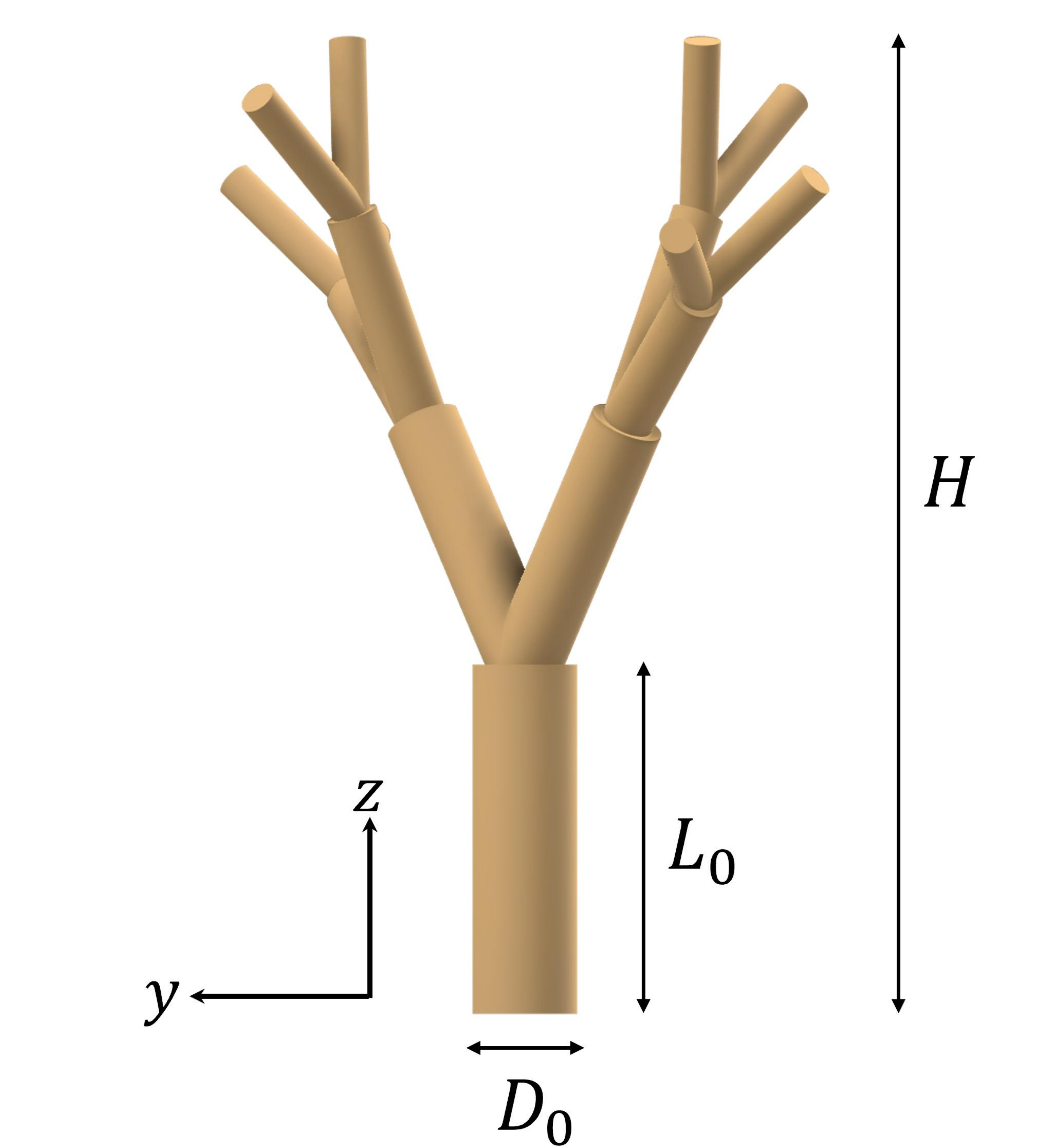}%
\hfill
\includegraphics[width=0.27\textwidth]{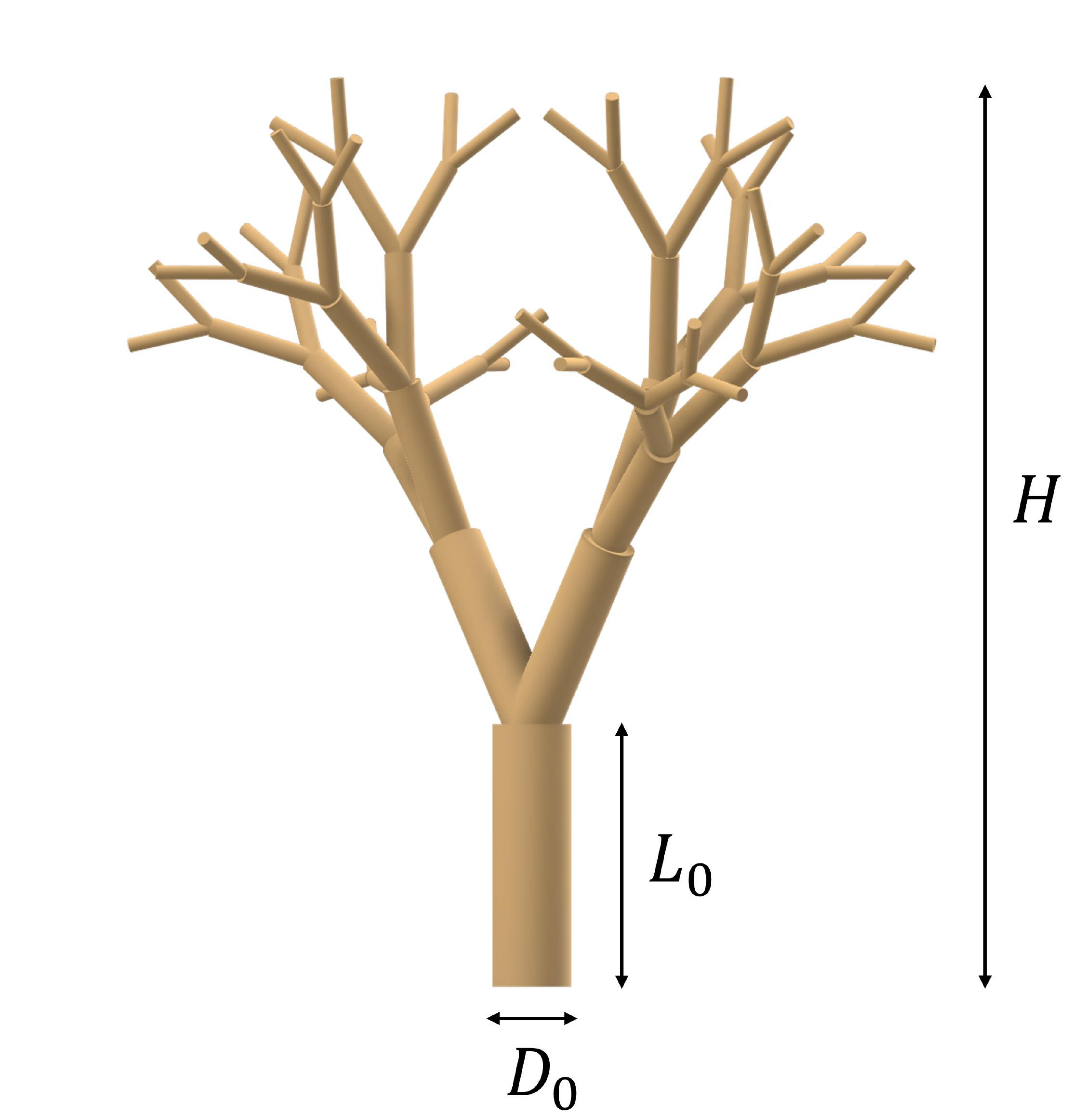}%
\hfill
\includegraphics[width=0.30\textwidth]{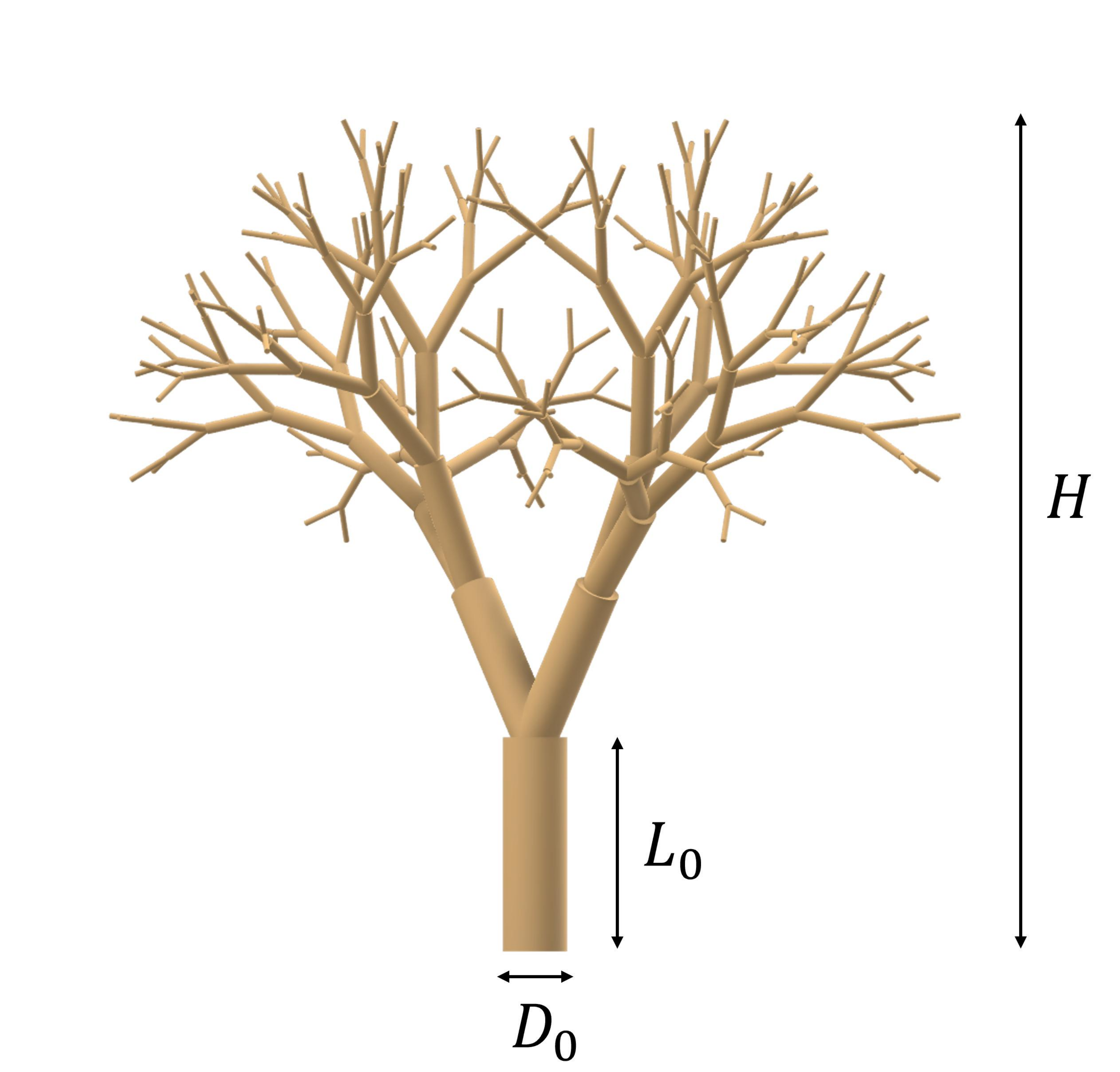}\\[1ex]
\includegraphics[width=0.27\textwidth]{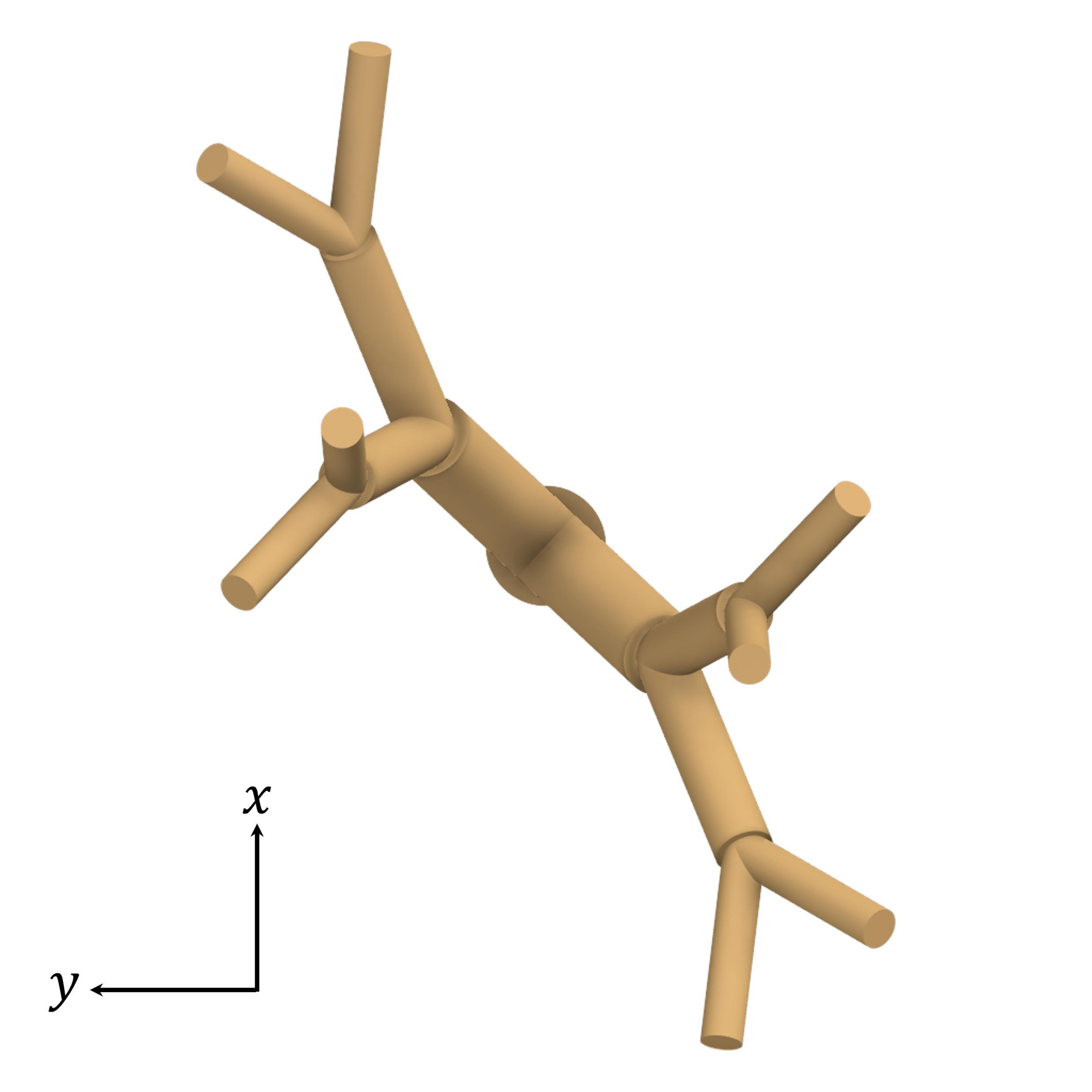}%
\hfill
\includegraphics[width=0.27\textwidth]{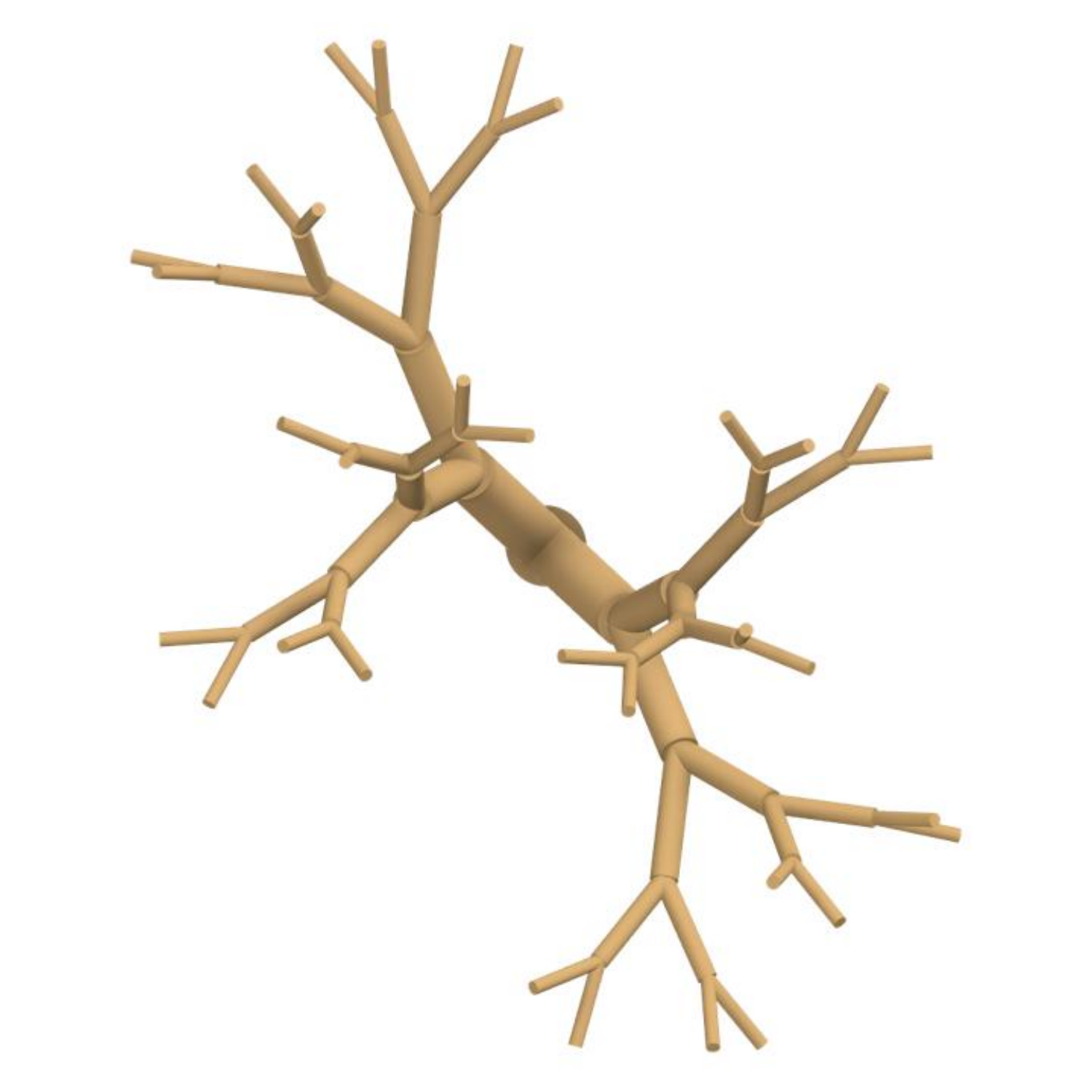}%
\hfill
\includegraphics[width=0.27\textwidth]{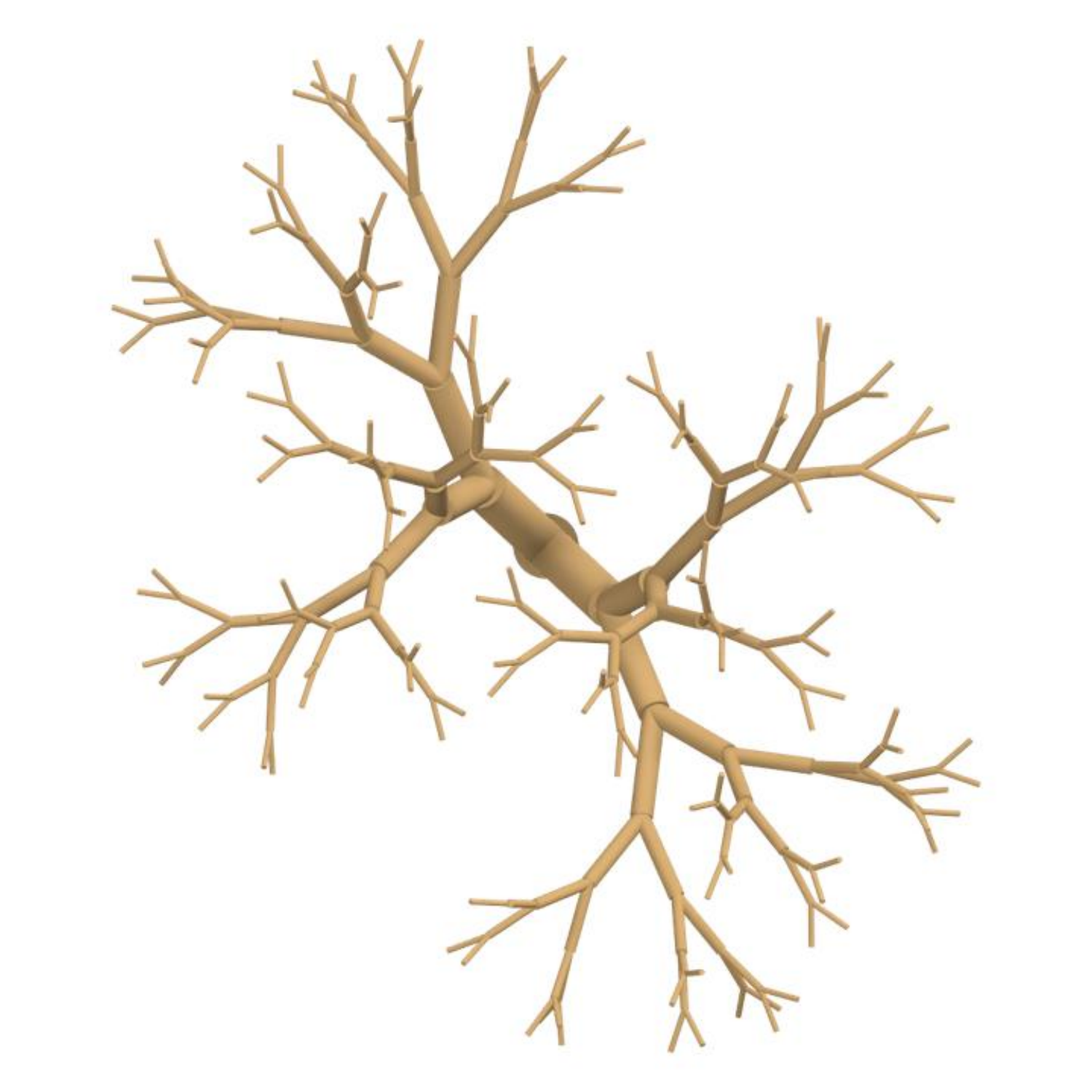}
\caption{\label{fig:tree_side_up}%
Fractal tree geometries. Top row: side views for (a) $n=4$, (b) $n=6$,
and (c) $n=8$. Bottom row: corresponding top views for (d) $n=4$, (e) $n=6$,
and (f) $n=8$. The parameter $n$ denotes the number of recursive branching
generations.}
\end{figure}

Three geometries of a single fractal tree type were generated using the
parametric L-system, based on the parameters listed in
Table~\ref{tab:L-system-parameters}, which follow those used in previous
work~\citep{Yin2025}.
Each tree was rescaled so that the total height was normalized to $1~\mathrm{m}$,
while maintaining the original horizontal and vertical proportions, as shown in
Fig.~\ref{fig:tree_side_up}. The fractal iteration numbers used were
$n=4$, $6$, and $8$. After normalization, the length and diameter of the
thickest branch---denoted $L_0$ and $D_0$, respectively---were adjusted
accordingly. The resulting geometric properties of each configuration are
summarized in Table~\ref{tab:tree-geometric-feature}.

Given scaling ratios of $0.8$ for both $r_1$ and $r_2$, branch lengths decrease
as $L_k=0.8 L_{k-1}$, where $k$ is the branching generation. Diameters scale
according to $D_k = 0.5^{0.5} D_{k-1}$, with a diameter ratio $q=0.5$. These
relationships define the hierarchical structure of the tree, from the trunk to
the finest branches.

The fractal dimension of the tree structures was quantified using a numerical
box-counting method~\citep{daSilva2006box}.
This approach involves overlaying the tree geometry with a grid of cubic boxes
of varying size $\delta$ and counting the number of boxes required to cover the
structure. The analysis of how this box count scales with the box size revealed
a consistent power-law relationship, yielding a fractal dimension of
approximately $1.77$ for the full tree structure~\citep{Yin2025}.

\begin{table}[htbp]
\caption{\label{tab:L-system-parameters}\centering
Parameters for the fractal trees shown in Fig.~\ref{fig:tree_side_up} using
Eq.~\eqref{eq:L-system-algorithm}. A minimum threshold $\min$ is imposed to
prevent unresolvable branching.}
\begin{tabular}{ccccccccccccc}
Model name & $r_1$ & $r_2$ & $\alpha_1$ & $\alpha_2$ & $\varphi_1$ &
$\varphi_2$ & $s_0$ & $w_0$ & $q$ & $e$ & $\min$ & $n$ \\
\hline
Basic $n=4$ & 0.8 & 0.8 & 30 & $-30$ & 137 & 137 & 100 & 30 & 0.5 & 0.5 & 0.02 & 4 \\
Basic $n=6$ & 0.8 & 0.8 & 30 & $-30$ & 137 & 137 & 100 & 30 & 0.5 & 0.5 & 0.02 & 6 \\
Basic $n=8$ & 0.8 & 0.8 & 30 & $-30$ & 137 & 137 & 100 & 30 & 0.5 & 0.5 & 0.02 & 8 \\
\end{tabular}
\end{table}

\begin{table}[htbp]
\centering
\caption{Geometric features of the fractal trees. Here $H$ is the overall tree height.}
\label{tab:tree-geometric-feature}
\begin{tabular}{lccc}
\hline
Model & Basic $n=4$ & Basic $n=6$ & Basic $n=8$\\
\hline
$D_0/H$ & 0.107 & 0.087 & 0.077\\
$L_0/H$ & 0.357 & 0.289 & 0.257\\
\hline
\end{tabular}
\end{table}

\subsection{Lattice Boltzmann method with adaptive mesh refinement}
\label{subsec:lbm}
Numerical simulations were carried out using a code based on LBM incorporating a cumulant collision term. This approach
has previously been applied successfully to simulate turbulent flows around
objects with complex geometries~\citep{Watanabe2021,Watanabe2023}.
The LBM, a weakly compressible approximation for solving incompressible flow
problems, is widely adopted for large-scale simulations of isothermal,
single-phase incompressible fluids. It models the fluid as an ensemble of
virtual particles and tracks the evolution of their velocity distribution
functions over time.

To accurately capture flows at high Reynolds numbers, we employed the D3Q27
cumulant collision model, \citep{Geier2015}
which offers excellent numerical stability and accuracy. This model minimizes
numerical viscosity when sufficient mesh resolution is used, thereby improving
the fidelity of the simulations. Additionally, the LBM is fully explicit and
does not require iterative procedures such as solving a Poisson equation for
pressure, as in traditional finite-difference methods. This feature contributes
to its high computational efficiency, especially for large-scale simulations.

To accurately resolve the boundary layer and wake regions around a fractal
tree, a high-resolution computational grid is essential. However, applying a
uniformly fine mesh throughout the entire domain would result in an
unmanageable number of grid points and demand excessive computational
resources. To balance accuracy with efficiency, we adopted an adaptive mesh
refinement (AMR) method based on an octree data structure. This approach allows
targeted refinement by recursively subdividing the computational grid, enabling
high-resolution meshes to be allocated selectively~\citep{Wahib2016}.

In our setup, the finest grid was assigned near the tree surface, where flow
gradients are most intense, while a moderately refined grid was applied to the
wake region. Coarser grids were used for regions farther from the tree to
minimize unnecessary computational cost. The simulation code executed most
fluid-dynamics calculations on GPUs, with the CPU responsible for mesh
generation and data output. LBM computations were carried out at the terminal
leaves of the octree, each of which contained an $8\times 8\times 8$ cell
block to optimize memory access for GPU operations. To further enhance
scalability, the domain was partitioned via a space-filling curve to evenly
distribute computational load and memory usage across multiple GPUs. Additional
details on the AMR–LBM implementation are available in~\citep{Watanabe2021}.

The LBM employs orthogonal grids for fluid computations. To accurately model
complex geometries such as the surface of a fractal tree, which does not align
with the grid, we avoided the conventional stair-stepped approximation.
Instead, we implemented a second-order accurate interpolated bounce-back
boundary condition~\citep{Bouzidi2001}
to more precisely define the object surface. Utilizing the D3Q27 lattice
configuration allowed us to capture surface orientations in 26 directions,
significantly improving the fidelity of the fractal tree representation.

Fluid forces on the object surface were computed using the momentum exchange
method~\citep{Wen2014}.
At each boundary grid point, the momentum change of the velocity distribution
function as it reflected off the surface was calculated. These local
contributions were then integrated over the entire object surface to determine
the net force acting on the object.

\subsection{Numerical setup}
\begin{figure}[htbp]
\centering
\includegraphics[width=0.9\textwidth]{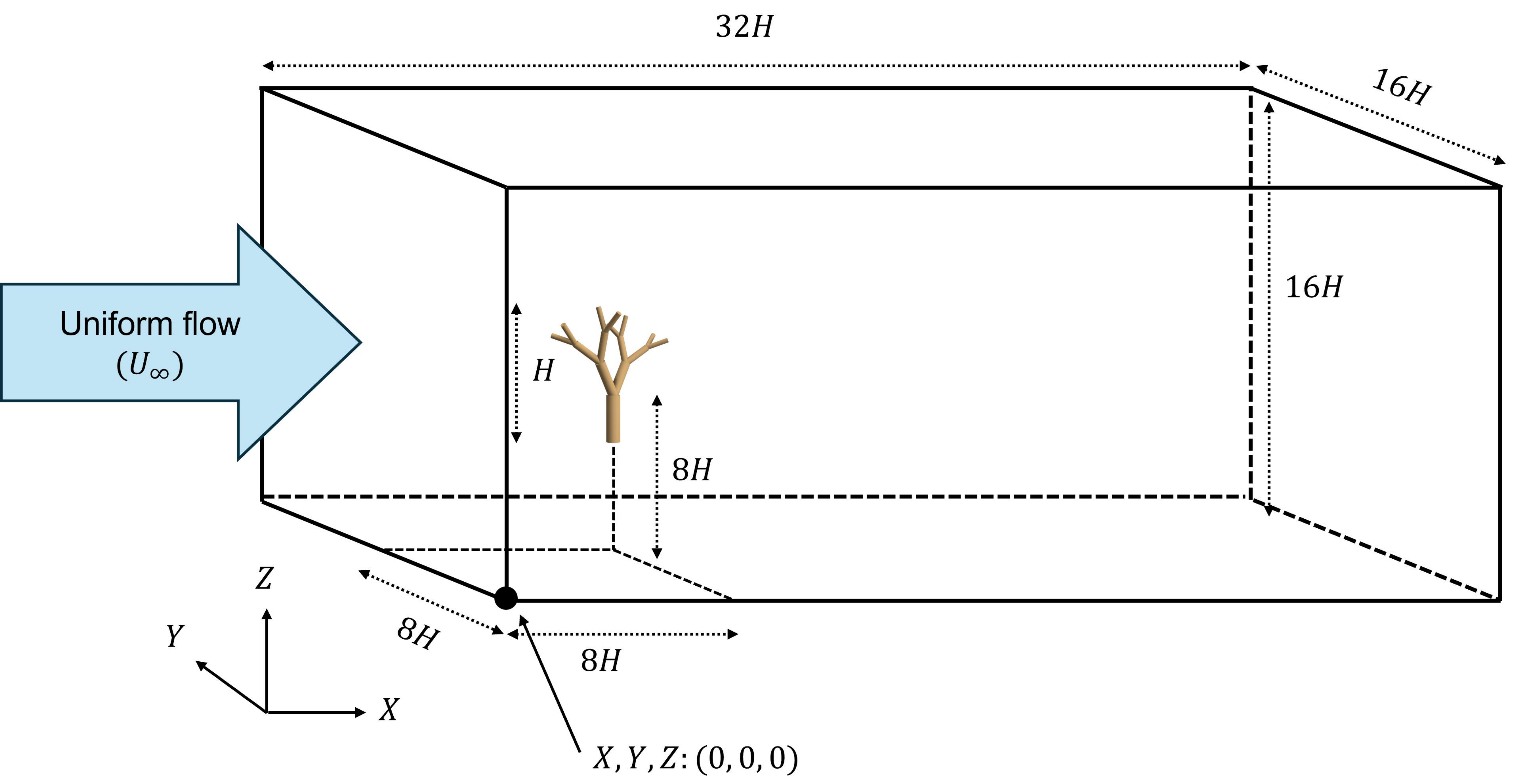}
\caption{Computational domain and arrangement of trees. $U_\infty$ represents
the uniform flow velocity in the $X$-direction of the inflow condition. The
$(X,Y,Z)$ coordinate system is used for the numerical simulations.}
\label{fig:simulation-setup}
\end{figure}

We adopted the same simulation setup as described in~\citep{Yin2025}, with the exception that the present study did not
include a detailed analysis of the wake region. The computational domain was
defined as $32H \times 16H \times 16H$. The coordinate system for the
numerical simulation was defined as $(X,Y,Z)$, as illustrated in
Fig.~\ref{fig:simulation-setup}. The center of mass of the tree was computed
and positioned at $(X,Y,Z) = (8H,8H,8H)$. A uniform inflow velocity $U_\infty$
in the $X$-direction was applied as the inflow condition. The trees were
modeled as rigid bodies, and their deformation due to fluid forces was
neglected.

Air properties at room temperature were used, with a density of
$\rho = 1.205~\mathrm{kg/m^3}$ and a kinematic viscosity of
$\nu = 1.512 \times 10^{-5}~\mathrm{m^2/s}$. Following~\citep{Yin2025}, the finest grid resolutions of
$\Delta x/H = 1/1024$, $1/1024$, and $1/2048$ were used near the tree surface
for Basic $n=4$, $6$, and $8$, respectively. Since a detailed wake-flow
analysis was not required in this study, we employed a coarser grid resolution
of $\Delta x/H = 1/256$ in the wake region, twice as coarse as that used in~\citep{Yin2025}.

\begin{figure}[htbp]
\centering
\includegraphics[width=0.90\textwidth]{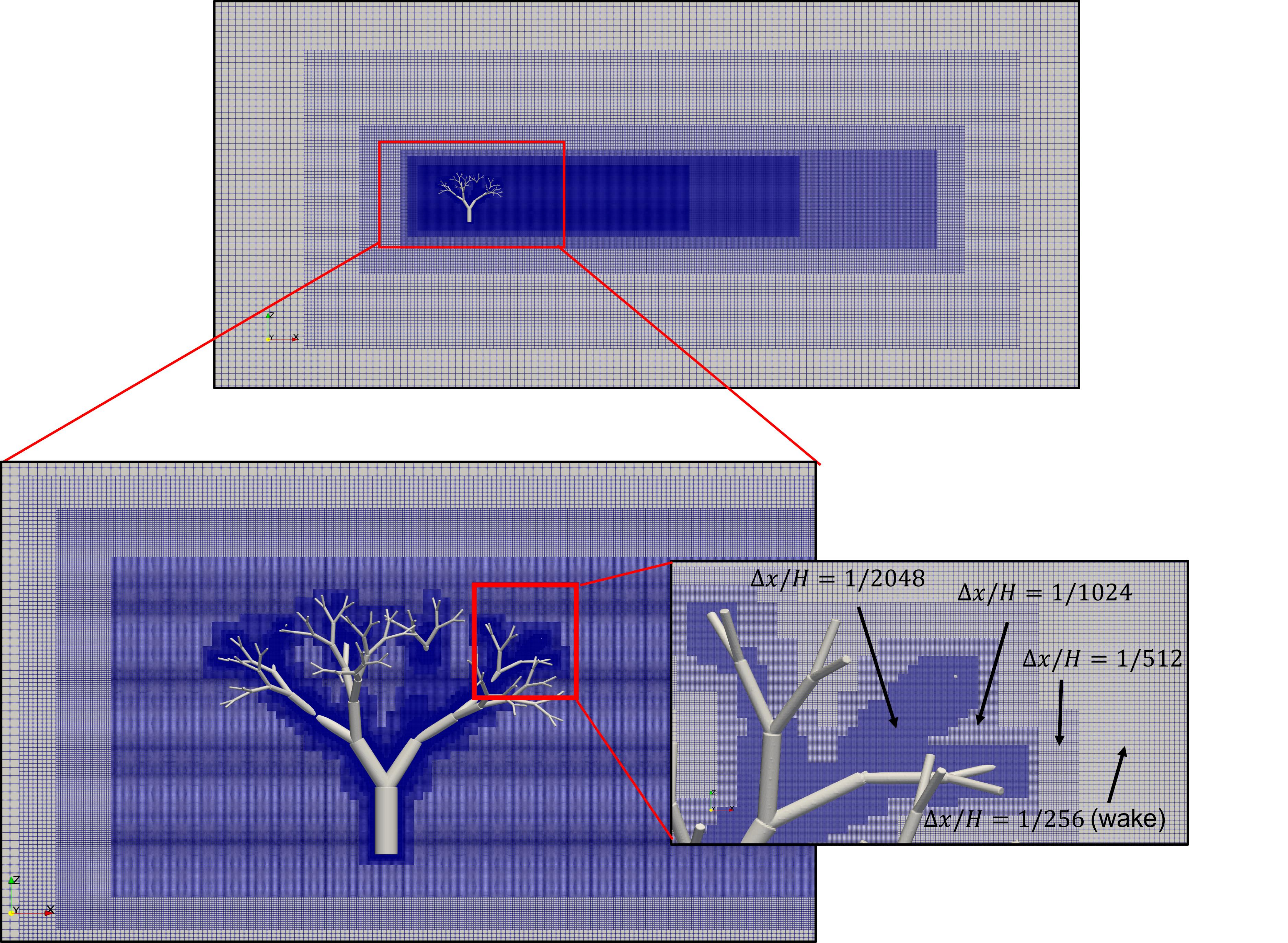}
\caption{Hierarchical computational mesh around the Basic $n=8$ tree model.
The mesh is refined progressively from the far field to the tree and wake
regions. The finest resolution, $\Delta x/H = 1/2048$, is applied near the
smallest branches, while coarser resolutions are used farther away and in the
wake region ($\Delta x/H = 1/256$).}
\label{fig:mesh-distribution}
\end{figure}

Figure~\ref{fig:mesh-distribution} shows a representative example of the AMR
grids for the case of Basic $n=8$. The total numbers of grid points were
$248{,}684{,}032$, $289{,}505{,}792$, and $901{,}405{,}696$ for Basic $n=4$,
$6$, and $8$, respectively. An outflow boundary condition was applied at the
downstream boundary of the domain, while inflow conditions were imposed on the
remaining boundaries. The lateral and vertical boundaries were positioned at a
distance of approximately $8H$ from the tree, ensuring negligible influence on
the flow field.

A no-slip condition was imposed on the tree surface using the interpolated
bounce-back method. Simulations were performed on the TSUBAME 4.0
supercomputer at Institute of Science Tokyo.

\subsection{Analytical framework}\label{sec:analytical-method}
In this study, we modeled the fractal trees as assemblies of cylindrical
segments with varying diameters and lengths. Leveraging this geometric
characterization, we performed an analytical estimate of the drag coefficients
of the tree. For each cylinder, the total drag force and the friction drag
force--denoted as $F_d$ and $F_f$, respectively--were calculated. The
estimated total drag force acting on the tree, $F_{D\text{-est}}$, and the
estimated friction drag force, $F_{F\text{-est}}$, were obtained by summing
these contributions across all segments. From these, the corresponding
analytical drag coefficients, namely the total drag coefficient
$C_{D\text{-est}}$ and the friction drag coefficient $C_{F\text{-est}}$, were
derived. Based on the standard drag decomposition $C_D = C_F + C_P$, the
analytical pressure drag coefficient was computed as
$C_{P\text{-est}} = C_{D\text{-est}} - C_{F\text{-est}}$.

The total and friction drag forces acting on each cylindrical segment were
estimated using
\begin{align}
F_d &= C_d \left(\frac{1}{2} \rho U_\infty^2 \right) (L D) \sin^3 \alpha,
\label{eq:total-force-cylinder}
\end{align}
\begin{align}
F_f &= C_f \left(\frac{1}{2} \rho U_\infty^2 \right) (L D) \sin^3 \alpha,
\label{eq:friction-force-cylinder}
\end{align}
where $C_d$ and $C_f$ are the total and friction drag coefficients of a
circular cylinder, $\alpha$ is the angle between the cylinder axis and the
flow direction, and $L$ and $D$ are the cylinder length and diameter. The
angular dependence, expressed as $\sin^3 \alpha$, follows the formulation
proposed in~\citep{kikai}.

The total drag coefficient $C_d$ was determined by fitting established
experimental data~\citep{panton2013incompressible,hoerner1965fluid}.
Since corresponding data for the friction drag coefficient $C_f$ are scarce, we
generated them using LBM simulations. The estimated drag coefficients of the
tree were then calculated as
\begin{align}
C_{D\text{-est}} &= \frac{F_{D\text{-est}}}{\frac{1}{2} \rho U_\infty^2 A_x},
\label{eq:estimate-total-drag-coeff}
\end{align}
\begin{align}
C_{F\text{-est}} &= \frac{F_{F\text{-est}}}{\frac{1}{2} \rho U_\infty^2 A_x},
\label{eq:estimate-friction-drag-coeff}
\end{align}
\begin{align}
C_{P\text{-est}} &= C_{D\text{-est}} - C_{F\text{-est}},
\label{eq:estimate-pressure-drag-coeff}
\end{align}
where $A_x$ denotes the projected frontal area of the tree in the streamwise
direction.

\section{Drag Dependencies: numerical simulation and analytical validation}

The total ($C_D$), friction ($C_F$), and pressure ($C_P$) drag coefficients were computed using the Lattice Boltzmann Method (LBM) over a range of $2{,}500\le Re_H \le 120{,}000$.

The total drag coefficient $C_D$ was computed directly from the total drag force $F_D$ [N] obtained in the LBM simulations:
\begin{align}
C_D = \frac{F_D}{\frac{1}{2} \rho U_\infty^2 A_x},
\label{eq:total_drag_coeff}
\end{align}
where $\rho$ [kg/m$^3$] is the density, $U_\infty$ [m/s] is the uniform inflow velocity, and $A_x$ [m$^2$] is the projected frontal area of the tree in the streamwise direction. The friction drag coefficient $C_F$ was evaluated by integrating the wall shear stress over the entire tree surface to find the total friction force $F_F$ [N], which was then made dimensionless:
\begin{align}
C_F = \frac{F_F}{\frac{1}{2} \rho U_\infty^2 A_x}.
\label{eq:friction_drag_coeff}
\end{align}
Finally, the pressure drag coefficient was derived as the difference between the total and friction components:
\begin{align}
C_P = C_D - C_F.
\label{eq:pressure_drag_coeff}
\end{align}

\begin{figure}[htbp]
\centering
\includegraphics[width=\linewidth]{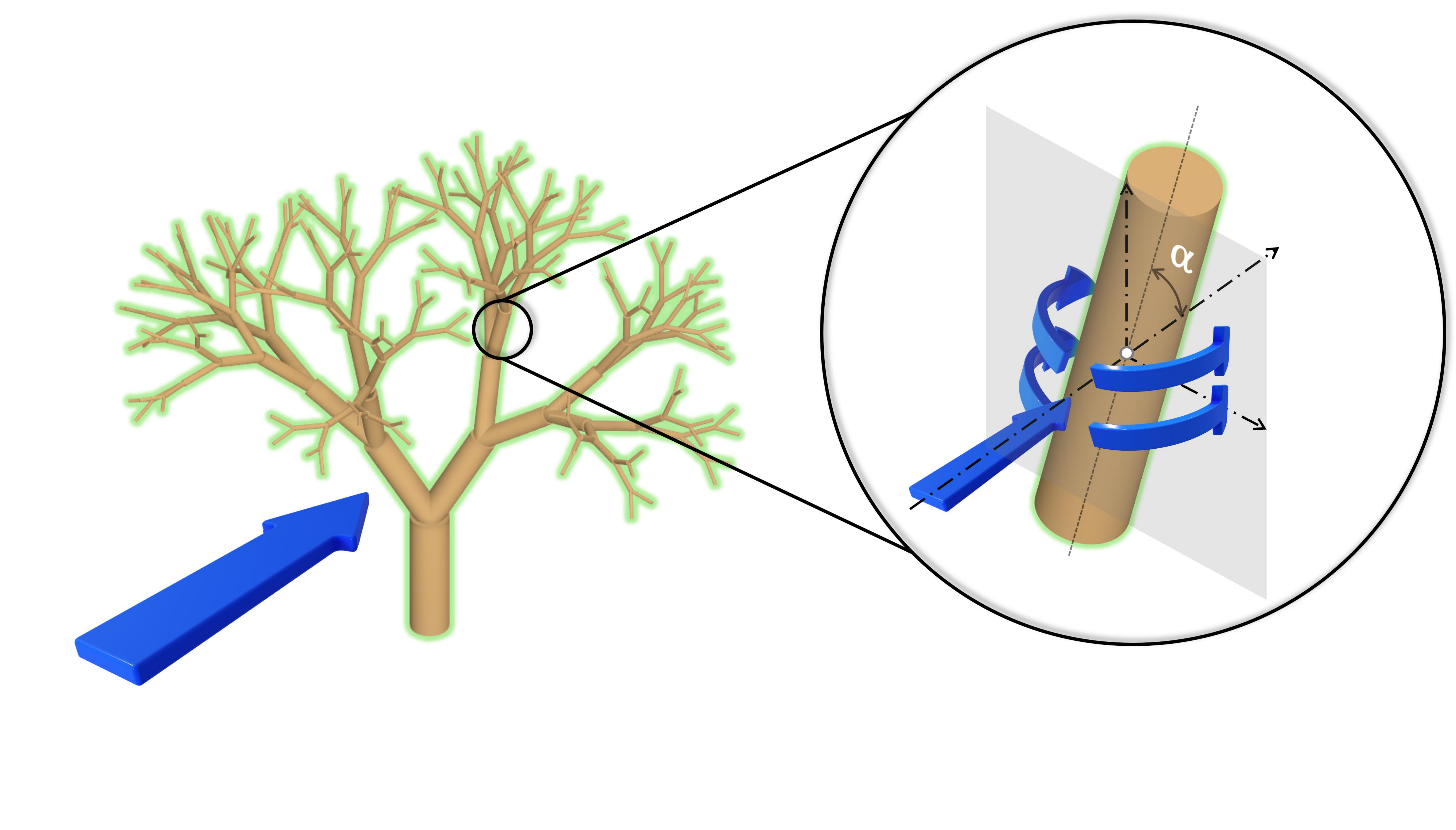}
\caption{Schematic of the analytical framework. The tree-like structure is
decomposed into a series of individual cylindrical segments. The orientation of
each segment is defined by the angle $\alpha$ between its central axis and the
streamwise direction of the uniform inflow velocity.}%
\label{fig:image-theoretical-way}
\end{figure}

Alongside these simulations, we use the analytical framework described in Sec.~\ref{sec:analytical-method}
to estimate the drag on tree-like structures. As illustrated schematically in
Fig.~\ref{fig:image-theoretical-way}, this approach decomposes the tree
into its constituent cylindrical elements; the drag on each branch segment is
evaluated using empirical relationships, and the total drag is obtained by
summing contributions across all segments.

\begin{figure}[htbp]
\centering
\begin{subfigure}{\linewidth}
\centering
\includegraphics[width=\linewidth]{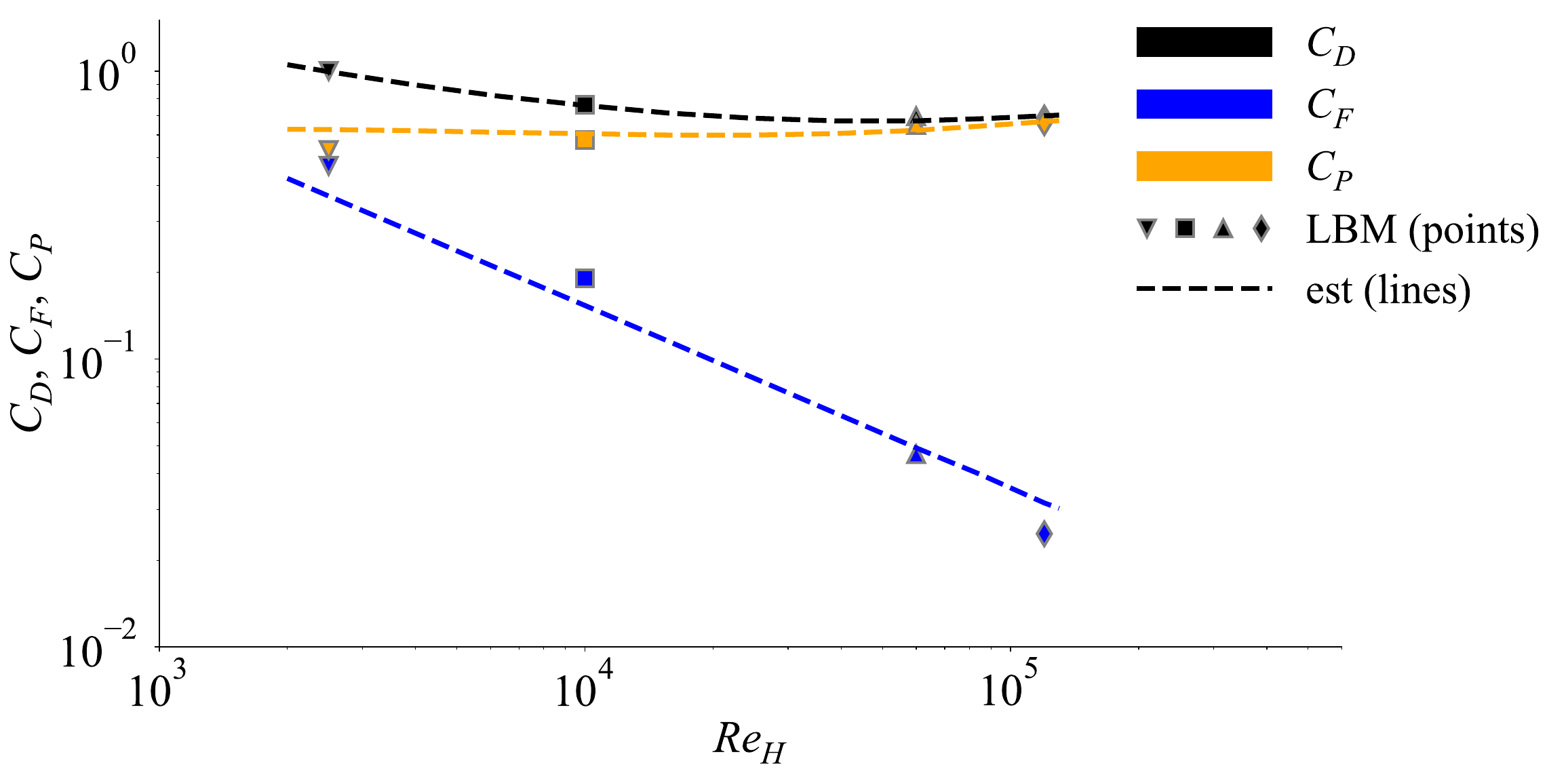}
\caption{Comparison against Reynolds number $Re_H$ for $n=6$.}
\label{fig:verify-Re}
\end{subfigure}

\begin{subfigure}{\linewidth}
\centering
\includegraphics[width=1.03\linewidth]{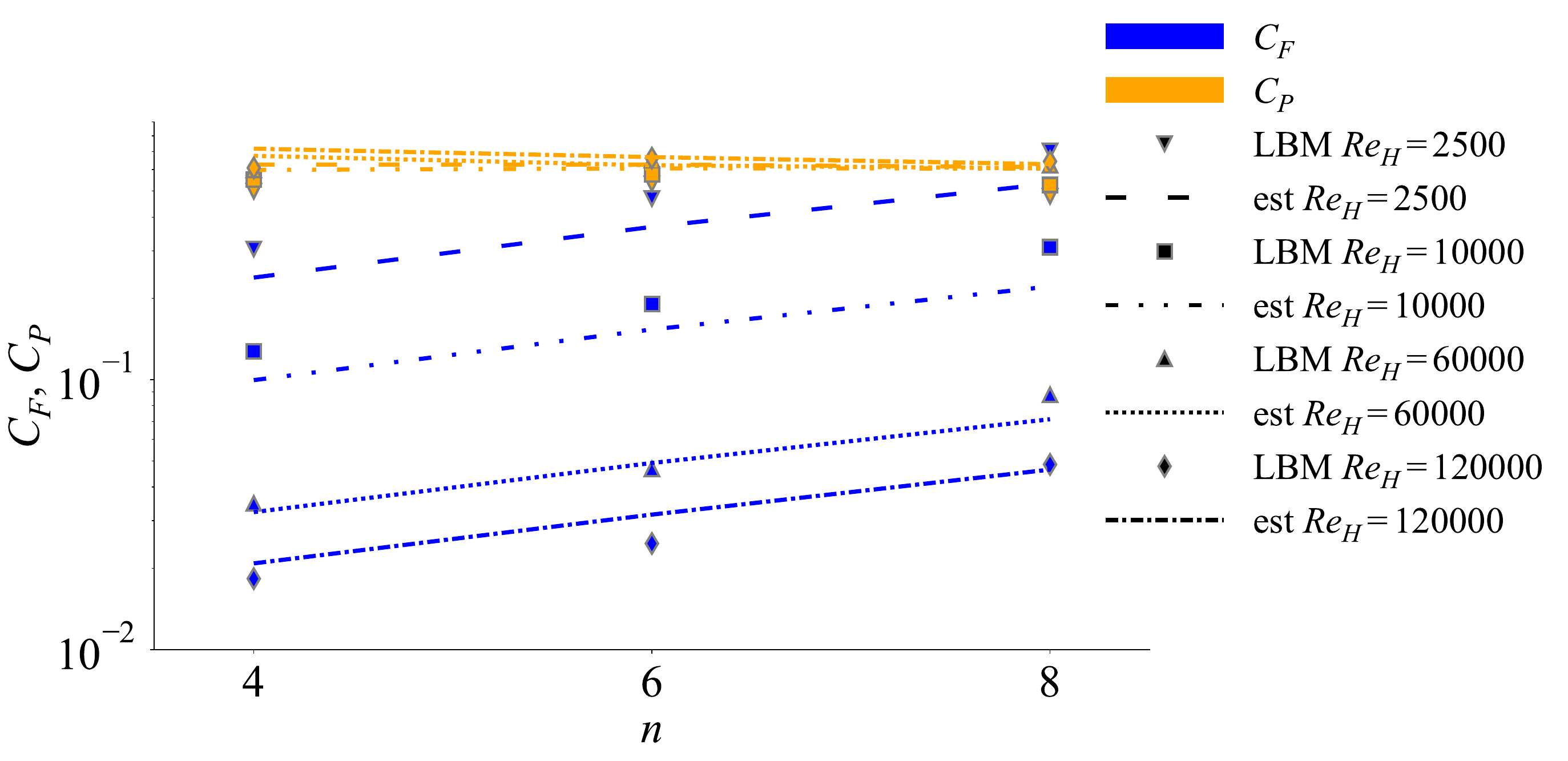}
\caption{Comparison against tree complexity $n$ for each Reynolds number.}
\label{fig:verify-n}
\end{subfigure}

\caption{Comparison of drag coefficients between LBM results (symbols) and
analytical estimates (lines). All LBM values are normalized by $C_D$ of LBM at
$n=6$, $Re_H = 2{,}500$, and all analytical values are normalized by the
corresponding analytical drag coefficient at the same condition, thereby
removing quantitative bias and highlighting qualitative agreement:
\textbf{(a)} comparison against Reynolds number $Re_H$ for $n=6$;
\textbf{(b)} comparison against tree complexity $n$ for each Reynolds number.}
\label{fig:validation}
\end{figure}

The results from both the LBM simulations and the analytical model are presented together in Fig.~\ref{fig:validation}. To assess the validity of the proposed analytical model, we compared its estimates with the LBM results, normalizing both datasets by their respective $C_D$ values at $n=6$ and $Re_H=2{,}500$ to facilitate a direct comparison of trends.

Focusing first on the LBM results (points in Fig.~\ref{fig:validation}), we observe the variation of $C_D$, $C_F$, and $C_P$ with $Re_H$ for the tree geometry with $n=6$ in Fig.~\ref{fig:verify-Re}. For this geometry, $C_D$ decreases monotonically with increasing $Re_H$ towards a nearly constant value. This trend is consistent with classical behavior observed in flows around isolated circular cylinders prior to the drag crisis~\citep{panton2013incompressible}, suggesting that similar aerodynamic mechanisms extend to more complex, branched structures. $C_F$ also decreases with increasing $Re_H$, but shows a marked drop beyond $Re_H \approx 60,000$, where $C_P$ reaches approximately 90\% of the total drag, becoming the dominant contributor. This behavior reflects the well-established fluid dynamic principle that the relative influence of viscous forces diminishes at higher Reynolds numbers, leading to pressure drag dominating the total drag~\citep{achenbach1968distribution,ohta2024friction}.

Figure~\ref{fig:verify-n} compares the LBM-computed $C_F$ and $C_P$ across tree geometries of increasing complexity at fixed $Re_H$. While $C_F$ increases monotonically with increasing $n$, $C_P$ remains approximately constant. This escalation of $C_F$ is likely due to the increasing proportion of thinner cylindrical components. As the iteration number $n$ increases, the tree becomes composed of a greater number of small branches, which locally experience lower Reynolds numbers. It is well known that friction drag increases relatively rapidly, whereas pressure drag varies only slightly as Reynolds number decreases in the low-Reynolds-number regime~\citep{White2006,Schlichting2017}. Therefore, $C_F$ of the fractal tree increases with increasing $n$, while $C_P$ remains nearly constant.

We now turn to the analytical estimates (lines in Fig.~\ref{fig:validation}). The raw analytical predictions show a systematic overestimation of the main drag components, with the total and pressure drag coefficients being overpredicted by around 40\%. This primary divergence is an expected consequence of the model's underlying assumptions, and it is noteworthy that the discrepancy in the friction drag coefficient is considerably smaller, averaging around 10\%. The analytical framework, by design, calculates drag on each cylindrical element independently and does not explicitly model wake shielding effects, where downstream branches are shielded by upstream ones, a phenomenon known to significantly reduce pressure drag~\citep{arie1983pressure,wang2020trapezoidal,zhou2024wake}. This mutual shielding can also be interpreted as a canopy-style sheltering (self-shading) effect.
Because the LBM--model discrepancy largely appears as a systematic offset in the pressure-drag component while the normalized $Re_H$ and $n$ trends are well captured,
such sheltering is expected to modify primarily the magnitude rather than the qualitative dependencies emphasized here.
The much closer agreement for friction drag supports this, indicating that viscous forces are less sensitive to such wake interactions.

Crucially, despite the quantitative offsets, the analytical model demonstrates excellent qualitative agreement with the LBM results. As shown in Fig.~\ref{fig:verify-Re} and~\ref{fig:verify-n}, it successfully captures the dependencies of the drag coefficients on both the Reynolds number ($Re_H$) and the tree complexity ($n$). The strong correspondence between the normalized trends confirms the analytical framework's utility as a computationally efficient and reliable tool for conducting parametric analyses and gaining qualitative insights into how structural parameters and flow conditions influence the drag on tree-like structures. Moreover, as the Reynolds number increases, the wake region tends to contract, leading to reduced interference among downstream elements. Consequently, the relative error introduced by neglected wake interactions is expected to diminish at higher $Re_H$. This supports the validity of extending the analytical framework to higher Reynolds number regimes examined in the following section.

\section{Drag crisis in fractal trees}

\subsection{Estimate and mechanism}
The drag crisis, an abrupt drop in drag coefficient associated with the transition of the boundary layer from laminar to turbulent flow, is a critical aerodynamic phenomenon for assessing tree stability and understanding urban airflow dynamics. However, direct observation of this phenomenon in tree-like structures has been limited due to experimental constraints and the prohibitively high computational cost of resolving flow at extreme Reynolds numbers.

To overcome these limitations, we employed the validated analytical framework introduced in the previous section to estimate the total drag coefficient over an extended Reynolds number range encompassing the regime where drag crisis is expected to occur. Figure~\ref{fig:drag-crisis} presents the estimated variation of total drag coefficient $C_D$ for each tree complexity up to $Re_H \sim 10^9$. This covers realistic scenarios. For instance, $Re_H$ between $10^5$ and $10^8$ corresponds to wind speeds from 1 m/s to 60 m/s acting on trees with heights between 1.5 m and 30 m.
The results suggest that drag crisis occurs near $Re_H \approx 3 \times 10^6$ for all geometries, similar in nature to the classical drag crisis observed for isolated circular cylinders~\citep{panton2013incompressible}.
Notably, the severity of the drag crisis, reflected in the steepness of the $C_D$ drop, diminishes with increasing $n$, i.e., as the tree geometry becomes more complex. This trend is attributed to the distribution of segment diameters within each tree. Since the Reynolds number associated with each cylindrical element is proportional to its diameter, only those with sufficiently large diameters experience the classic drag crisis. In models with larger $n$, the number of small-diameter segments increases, many of which remain below the critical Reynolds number. As a result, a greater proportion of the structure maintains pre-crisis drag characteristics, softening the overall transition and producing a more gradual slope in the $C_D$ curve.

\subsection{Influence of inflow turbulent intensity}
Real urban winds are characterized by non-negligible inflow turbulence intensity.
For an isolated circular cylinder, increasing inflow turbulence promotes earlier
boundary-layer transition: the onset of the drag crisis shifts to lower Reynolds
numbers and the drag reduction becomes more gradual~\citep{Yao2019InflowTurbulenceCylinder}.
Here the turbulence intensity is defined as $I_u = u'_{\mathrm{rms}}/U_\infty$, where
$u'_{\mathrm{rms}}$ is the root-mean-square of the streamwise velocity fluctuations
and $U_\infty$ is the mean inflow speed.
In the experiments of \citet{Yao2019InflowTurbulenceCylinder}, the inflow
turbulence intensity is approximately $I_u \approx 8\%$, which is broadly
representative of near-neutral atmospheric boundary-layer flows over rough
urban surfaces.

\begin{figure}[htbp]
\centering
\includegraphics[width=1.0\linewidth]{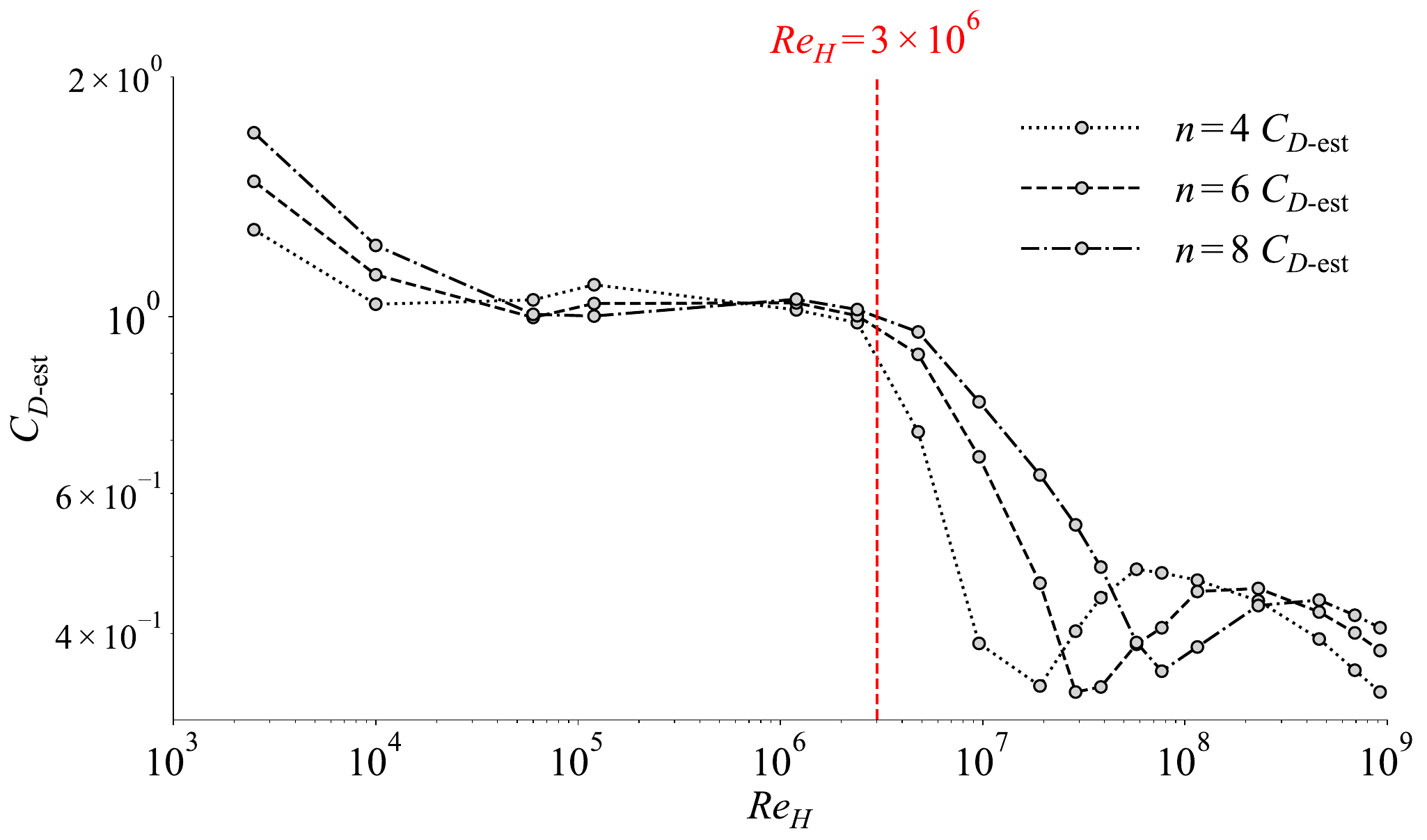}
\caption{Estimated total drag coefficients ($C_{D\text{-est}}$) against the
Reynolds number $Re_H$ for each tree complexity. In all tree complexities, the
drag crisis (i.e., the sharp drop of total drag associated with the transition
from laminar to turbulent flow) is estimated to occur around
$Re_H \approx 3\times 10^{6}$. The gradients become gentler as $n$ increases.}%
\label{fig:drag-crisis}
\end{figure}

\begin{figure}[htbp]
\centering
\includegraphics[width=1.0\linewidth]{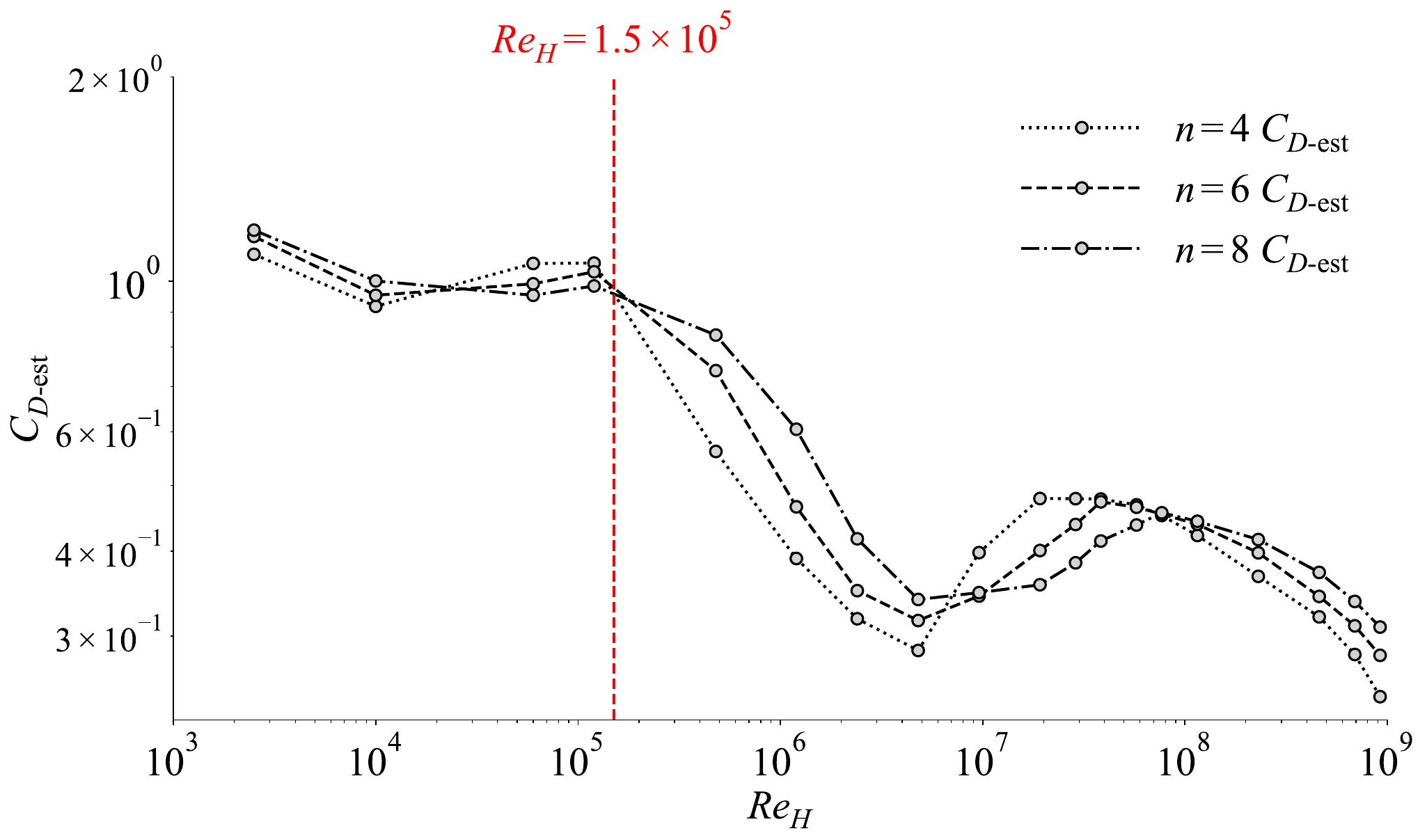}
\caption{Sensitivity of the estimated drag-crisis curve to turbulent inflow. The element-wise drag
law was modified to reflect the reported effect of inflow turbulence, characterized by a streamwise
turbulence intensity $I_u \approx 8\%$ for circular cylinders (ratio of root-mean-square velocity fluctuations
to the mean velocity): the crisis onset shifts to lower Reynolds number and the drag reduction
becomes more gradual. Such a turbulence intensity is broadly representative of near-neutral
atmospheric boundary-layer flows over rough urban surfaces. The vertical dashed line indicates
the shifted onset Reynolds number adopted in this sensitivity test ($Re_{H,cr} = 1.5\times10^5$).
}
\label{fig:drag-crisis-Iu8}
\end{figure}

To assess how such effects may influence the tree-scale prediction, we conducted
a sensitivity test by embedding this turbulent-inflow cylinder trend into the
element-wise drag law used in our branch-wise model.
Specifically, we reconstructed the element-wise drag curve in the crisis range
using the turbulent-inflow cylinder $C_D(Re)$ data reported by
\citet{Yao2019InflowTurbulenceCylinder}.
Figure~\ref{fig:drag-crisis-Iu8} shows the resulting estimated total drag
coefficient curves.
Relative to the uniform-inflow estimate in Fig.~\ref{fig:drag-crisis}, the onset
of the crisis shifts to lower $Re_H$ and the drop in $C_{D\text{-est}}$ becomes
more gradual, in line with the underlying cylinder behavior.

Importantly, however, the key structural trend remains unchanged:
increasing the complexity $n$ still moderates the overall crisis, producing a
progressively smoother decline of $C_{D\text{-est}}$.
Interpreting the turbulence-modified onset Reynolds number in terms of full-scale
parameters further highlights the relevance of this regime.
Using $Re_H = U_\infty H/\nu$ with $Re_{H,\mathrm{cr}} \approx 1.5\times 10^{5}$
(from Fig.~\ref{fig:drag-crisis-Iu8}), a $10~\mathrm{m}$ tree in air
($\nu \approx 1.5\times 10^{-5}~\mathrm{m^2/s}$) would reach the crisis threshold
at a mean wind speed of only $U_\infty \approx 0.2~\mathrm{m/s}$.
Thus, typical near-surface winds of order $1$--$10~\mathrm{m/s}$ acting on
trees of order $10$--$30~\mathrm{m}$ height correspond to $Re_H$ in the range
$10^{6}$--$10^{8}$, i.e., within or beyond the drag-crisis regime.

The present model represents only the woody structure and does not explicitly
resolve foliage-scale processes.
Foliage can affect the absolute drag magnitude, particularly at low to moderate
wind speeds; however, under strong winds, measurements indicate that woody stems
and branches still contribute a substantial fraction of the total load, while
leaf reconfiguration and damage or abscission can cause the incremental drag
contribution of leaves to saturate at high wind speeds~\citep{Vollsinger2005CrownStreamlining,Li2021CrownHighWinds}.
Taken together with the turbulence sensitivity test, these considerations
suggest that the present estimates provide a useful baseline: realistic factors
may shift the onset and magnitude of the crisis, yet the dependence on $n$ and
the qualitative high-$Re_H$ tendency remain robust.
In this sense, the framework offers a practical predictor for exploring how
branching complexity shapes drag-crisis behavior, while more detailed
representations of foliage and reconfiguration can be incorporated in future
work to improve quantitative accuracy.

\section{Implications for Drag Reversal and Urban Tree Management}

The moderation of the drag crisis across more complex tree structures has important aerodynamic implications. In particular, the gradual transition in $C_D$ can lead to a reversal in the relative magnitude of drag coefficients among different geometries at certain Reynolds numbers. For instance, while more complex trees (larger $n$) generally show larger drag in the subcritical $Re_H$ regime ($Re_H \sim 10^7$), they may exhibit smaller total drag than simpler geometries in the supercritical regime ($Re_H \sim 10^8$) due to their muted drops in $C_D$. In other words, reducing structural complexity does not necessarily lower drag across all flow conditions.

This finding has practical implications for urban tree management. Pruning reduces structural complexity by removing smaller branches, effectively lowering $n$ and increasing the relative proportion of large-diameter segments, those more likely to experience a full drag crisis. Under strong winds (e.g., $Re_H\sim10^8$, corresponding to 20-30 m tall trees and 40-60 m/s winds), such trees may experience greater aerodynamic loads, elevating the risk of breakage for certain conditions. These results challenge the conventional assumption that pruning reliably reduces aerodynamic loading~\citep{smiley2006pruning,cao2012wind}. Our results thus offer a scientific basis for reassessing the aerodynamic consequences of tree pruning and for optimizing maintenance strategies to enhance urban safety.

\section{Conclusions and outlook}
We investigated the aerodynamic drag of fractal tree-like structures over a wide Reynolds-number range by combining large-scale lattice Boltzmann (LBM) simulations with a reduced-order, branch-wise analytical model.
The LBM results quantified how the total drag coefficient and its friction/pressure decomposition depend on the tree-height-based Reynolds number $Re_H$ and the branching complexity $n$.
Although the analytical model exhibits a systematic offset in magnitude, primarily attributable to unresolved mutual shielding among branches, it reproduces the normalized $Re_H$ and $n$ trends observed in LBM.
This supports its use as a computationally efficient baseline framework for exploring the aerodynamic consequences of tree architecture.

Using this framework together with established element-level drag behavior, we estimated tree-scale drag-crisis characteristics up to $Re_H\sim10^{9}$.
The estimates consistently indicate that the overall drag-crisis drop becomes progressively gentler as $n$ increases, because an increasing proportion of elements remains below the critical Reynolds number associated with the crisis.
A sensitivity test based on turbulent-inflow cylinder data with a streamwise turbulence intensity $I_u\approx 8\%$, representative of near-neutral atmospheric-boundary-layer flows over rough urban surfaces, shifts and smooths the crisis curve. However, it preserves the key $n$-dependence (Fig.~\ref{fig:drag-crisis-Iu8}), suggesting robustness to realistic inflow conditions.
Interpreting the turbulence-modified onset Reynolds number $Re_{H,\mathrm{cr}}\approx 1.5\times10^{5}$ in terms of full-scale parameters further highlights the practical importance of this regime: for a $10~\mathrm{m}$ tree in air, $Re_{H,\mathrm{cr}}$ corresponds to a mean wind speed of only $U_\infty\approx0.2~\mathrm{m/s}$.
Consequently, typical atmospheric boundary-layer winds of order $1$--$10~\mathrm{m/s}$ acting on trees $10$--$30~\mathrm{m}$ tall place the flow around the tree well within or beyond the drag-crisis regime.
While additional processes such as foliage, reconfiguration, and sheltering can modify the absolute drag level, the present results provide a useful baseline that captures the qualitative dependence of drag on $Re_H$ and structural complexity in the high-$Re$ regime.

These findings further imply that the relative drag among different tree geometries can reverse across Reynolds-number regimes: more complex structures may exhibit larger drag in subcritical conditions, yet smaller drag than simpler geometries in supercritical conditions due to their muted drag-crisis drop.
Accordingly, reducing structural complexity does not necessarily reduce aerodynamic loading under all wind conditions.
This has practical implications for urban tree management, as pruning that removes smaller branches can increase the relative proportion of large-diameter elements that undergo a pronounced drag crisis and may elevate peak loads under strong winds.
Therefore, the present baseline framework offers a scientific basis for reassessing aerodynamic consequences of pruning and for informing maintenance strategies aimed at urban safety.

Future work will incorporate foliage and reconfiguration/fluid--structure interaction and will further validate the extrapolation toward higher-Re conditions, aiming at improved quantitative prediction under realistic urban winds.

\bibliographystyle{elsarticle-harv}
\bibliography{cas-refs}

\end{document}